\documentclass[11pt,aps,nofootinbib,prd,aps,epsf,floats,axodraw,amsmath,amssymb,amsfonts]{revtex4}
\usepackage{amsmath, amssymb}
\newcommand{\mathsym}[1]{{}}

\usepackage{graphicx}
\usepackage{amsmath}
\usepackage{amssymb}
\usepackage{bm}
\newcommand{\be}{\begin{equation}}
\newcommand{\ee}{\end{equation}}
\newcommand{\bea}{\begin{eqnarray}}
\newcommand{\eea}{\end{eqnarray}}

\newcommand{\rem}[1]{}
\newsavebox{\PSLASH}
 \sbox{\PSLASH}{$p$\hspace{-1.8mm}/}
 
\renewcommand{\theequation}{\thesection.\arabic{equation}}
\newcounter{saveeqn}
\newcommand{\add}{\addtocounter{equation}{1}}
\newcommand{\alpheqn}{\setcounter{saveeqn}{\value{equation}}%
\setcounter{equation}{0}%
\renewcommand{\theequation}{\mbox{\thesection.\arabic{saveeqn}{\alph{equation}}}}}
\newcommand{\reseteqn}{\setcounter{equation}{\value{saveeqn}}%
\renewcommand{\theequation}{\thesection.\arabic{equation}}}

 \newsavebox{\notrightarrow}
 \sbox{\notrightarrow}{$\to$\hspace{-4mm}/}
 
 \newsavebox{\PARTIALSLASH}
 \sbox{\PARTIALSLASH}{$\partial$\hspace{-1.6mm}/}
 
 \newsavebox{\ASLASH}
 \sbox{\ASLASH}{$A$\hspace{-2.1mm}/}
 
 \newsavebox{\KSLASH}
 \sbox{\KSLASH}{$k$\hspace{-1.8mm}/}
 
 \newsavebox{\LSLASH}
 \sbox{\LSLASH}{$\ell$\hspace{-1.8mm}/}
 
 \newsavebox{\QSLASH}
 \sbox{\QSLASH}{$q$\hspace{-1.8mm}/}
 
 \newsavebox{\DSLASH}
 \sbox{\DSLASH}{$D$\hspace{-2.2mm}/}
 
 \newsavebox{\DbfSLASH}
 \sbox{\DbfSLASH}{${\mathbf D}$\hspace{-2.8mm}/}
 
 \newsavebox{\DELVECRIGHT}
 \sbox{\DELVECRIGHT}{$\stackrel{\rightarrow}{\partial}$}
 
 \newcommand{\blue}{\IfColor{\textCadetBlue}{}}
\newcommand{\black}{\IfColor{\textBlack}{}}
\newcommand{\red}{\IfColor{\textRed}{}}
\newcommand{\green}{\IfColor{\textOliveGreen}{}}
\newcommand{\lila}{\IfColor{\textRedViolet}{}}

\begin{document}
\begin{flushright}
 [math-ph]
\end{flushright}
\title{$\star$-Cohomology, Third Type Chern Character and Anomalies in\\General Translation-Invariant Noncommutative Yang-Mills}

\author{Amir Abbass Varshovi}\email{ab.varshovi@sci.ui.ac.ir/amirabbassv@ipm.ir/amirabbassv@gmail.com}

\affiliation{Faculty of Mathematics and Statistics, Department of Applied Mathematics
and Computer Science, University of Isfahan, Isfahan, IRAN.\\
School of Mathematics, Institute for Research in Fundamental
Sciences (IPM), P.O. Box: 19395-5746, Tehran, IRAN.}
\begin{abstract}
       \textbf{Abstract\textbf{:}} A representation of general translation-invariant star products $\star$ in the algebra of $\mathbb{M}(\mathbb{C})=\lim_{N \to \infty} \mathbb{M}_N(\mathbb{C})$ is introduced which results in the Moyal-Weyl-Wigner quantization. It provides a matrix model for general translation-invariant noncommutative quantum field theories in terms of the noncommutative calculus on differential graded algebras. Upon this machinery a cohomology theory, the so called $\star$-cohomology, with groups $H_\star^k(\mathbb{C})$, $k\geq 0$, is worked out which provides a cohomological framework to formulate general translation-invariant noncommutative quantum field theories based on the achievements for the commutative fields, and is comparable to the Seiberg-Witten map for the Moyal case. Employing the Chern-Weil theory via the integral classes of $H_\star^k(\mathbb{Z})$ a noncommutative version of the Chern character is defined as an equivariant form which contains topological information about the corresponding translation-invariant noncommutative Yang-Mills theory. Thereby we study the mentioned Yang-Mills theories with three types of actions of the gauge fields on the spinors, the ordinary, the inverse, and the adjoint action, and then some exact solutions for their anomalous behaviors are worked out via employing the homotopic correlation on the integral classes of $\star$-cohomology. Finally, the corresponding consistent anomalies are also derived from this topological Chern character in the $\star$-cohomology. \\
       
\noindent \textbf{Keywords\textbf{:}} General Translation-Invariant Noncommutative Star Product, $\star$-Cohomology, Translation-Invariant Noncommutative Yang-Mills Theory, Third Type Chern Character, Abelian Anomaly, Consistent Anomaly, BRST.
\end{abstract}

\pacs{} \maketitle


\section{Introduction}\label{introduction}

\par Translation-invariant noncommutative star products $\star$, as the algebraic extension of the Moyal product, has been extensively considered in theoretical physics since the beginning of the current century, when it was broadly believed that a modification of the Moyal formulation will improve the UV/IR mixing problem in the perturbative renormalization of noncommutative quantum field theories.\footnote{See for example the pioneering works in this topic \cite {kossow, krajewski, woronowicz, tanasa1, tanasa2, tanasa-vitale, lizzi, meljanac, galluccio}. For a more complete list of such references see \cite{galluccio} and \cite{varshovi0} and the references therein.} However, the most important development of such generalizations, the so called Wick-Voros star product, goes back to three decades ago in \cite {voros}, wherein the author studied the WKB approximation via the Bargman representation with no address to its application in the theory of quantum fields.\footnote{Although, the Wick-Voros star product was introduced in 1989 \cite {voros}, but the discovery of its significance for quantum fields returns to about twenty years after its inception. See for example \cite {kossow, lizzi}.}

\par The generalization of the Moyal formula to translation-invariant noncommutative star products was then studied in the setting of a cohomology theory, the so called $\alpha^*$-cohomology \cite{lizzi, varshovi1, varshovi2}, where it was cleared that all quantum behaviors and the perturbative/non-perturbative properties of a quantum field theory (such as the scattering matrix, the renormalizability, the UV/IR mixing pathologies and the Green's functions structures) are invariant through the $\alpha^*$-cohomology classes \cite {varshovi2}. Moreover, it was discovered that for any general translation-invariant star product there exists a cohomologous Moyal product, known as the harmonic 2-cocycle \cite{varshovi1}. Therefore, the renormalizability of a noncommutative quantum field theory with a general translation-invariant noncommutative star product $\star$ is understood to be equivalent to the renormalizability of that theory with the Moyal multiplication formula.

\par Although the perturbative features, such as the renormalizability and the anomalous behaviors of quantum field theories with general translation-invariant noncommutative star products $\star$ were studied extensively in the last decade,\footnote{For example see \cite{ardalan2, chelabi, karim, ghasemkhani-varshovi, lizzi2, lizzi3, samary, tanasa1, tanasa2, tanasa-vitale, lizzi, meljanac, galluccio} and the references therein.} but, nonetheless, the geometric aspects and the topological properties of such theories are still beeing unclear. This topic on the one hand is intimately correlated to the fascinating achievements of noncommutative geometry \cite{connes-book}, wherein the algebra of noncommutative quantum fields is considered as the only fundamental structure, and on the other hand, it is strongly attached to the geometry and the topology of the underlying smooth manifolds, e.g. the spacetime and the corresponding fiber bundles \cite{nakahara}.

\par Anomalies are in fact one the most familiar topics which reveal significant information of the mentioned structures of the corresponding theory. Perturbative calculations in translation-invariant noncommutative Yang-Mills theoires lead to solutions for anomalies which are described by gauge fields, their derivations, and their multiplication with star product, known as the noncommutative Chern character.\footnote{See for example \cite{ardalan2, varshovi-consist} and the references therein. Particularly, \cite{ardalan2} applies the perturbative method and \cite{varshovi-consist} uses a non-perturbative approach.} On the other hand, the noncommutative geometric description of the geometry and topology of anomalies in general noncommutative gauge theories, including the Yang-Mills and the theory of general relativity, is given in terms of the cyclic (co)homology groups and the corresponding Connes-Chern characters \cite{perrot, perrot2}.

\par However, it has not been rigorously cleared yet how these formulations are correlated. In fact, it is absolutely unknown if the perturbative solution for translation-invariant noncommutative anomaly defines an integral de Rham cohomology class of the spacetime. Moreover, if it is so, we do not know to which cohomology class it belongs. It is also ambiguous how its cohomology class depends on the gauge field $A_\mu$ as a topological invariant. On the other hand, there is no definite derivation of the noncommutative Chern character via some cohomological framework, and consequently the topological meaning of the Abelian anomaly in noncommutative Yang-Mills theories with translation-invariant noncommutative star products is totally based on the intuitions rather than the firm detailed proofs.

\par This is while noncommutative geometry has not have a clear de Rham cohomology description for noncommutative algebras, such as that of the noncommutative fields with general translation-invariant noncommutative star products \cite {connes-book, khalkhali}. In addition, there are smooth manifolds, the spacetime and the Yang-Mills principal bundle, in the background of the corresponding noncommutative theory and we expect that the mentioned topological aspects, e.g. the noncommutative Chern character, to be described in terms of their characteristic classes in the de Rham cohomology groups.

\par To answer the above problems a new cohomology theory, the so called $\star$-cohomology, with groups $H_\star^k(X,\mathbb{C})$, $k\geq 0$, over the spacetime manifold $X$ is defined which classifies the operator valued differential forms on the commutative part of $X$ with extra dimensions on its noncommutative part. This operators are in fact elements of $\mathcal{L}(H)$ which define a representation of general translation-invariant noncommutative star product $\star$, while here $H$ is a separable Hilbert space. Thus, its classes are in principal represented by the elements of commutative geometry, whereas they contain an intrinsic noncommutativity due to $\star$. Actually, the $\star$-cohomology groups have intimate correlations to those of both the de Rham and the cyclic (co)homologies, and thus can be considered as the cohomological formulation of the Seiberg-Witten map \cite {witten} for general translation-invariant noncommutative star products.

\par Due to the commutative geometric background of the $\star$-cohomology, the Chern-Weil theory is employed and three types of the Chern character are derived accordingly. We show that the third type Chern character represents an integral class of the $\star$-cohomology and defines the Abelian anomaly, i.e. the noncommutative Chern character, in translation-invariant noncommutative Yang-Mills. Applying some homotopy formulations in the integral classes of $H_\star^{k}(X,\mathbb{Z})$, we establish its topological invariance and characterize its class in terms of the de Rham cohomology. This describes thoroughly the geometry and the topology of the corresponding theory and enables us to calculate the topological index of the noncommutative anomaly for any general translation-invariant noncommutative star product.

\par Three types of translation-invariant noncommutative Yang-Mills theories with the ordinary (type A), the inverse (type B), and the adjoint (type C) actions of the gauge fields on the spinors are studied by means of the $\star$-cohomology formulations. The corresponding chiral Abelian anomalies of these theories are extracted by applying a homotopy formula for the $\star$-cohomology class of the third type Chern character. It is shown that the type C translation-invariant noncommutative Yang-Mills is anomaly free for the chiral current.

\par Moreover, the BRST formulation is also worked out in the framework of the $\star$-cohomology and thereby the noncommutative versions of the Chern-Simons form, the consistent anomaly, and the consistent Schwinger term for the mentioned types A, B, and C are derived by trnasgression in the $\star$-cohomology classes due to the third type Chern character. Finally, the corresponding $\star$-BRST cohomology classes are extracted with some rather complicated formulas which belong to the integral classes of the $\star$-cohomology of the noncommutative gauge transformation group.


\par
\section{Translation-Invariant Star Product and $\star$-Cohomology}
\setcounter{equation}{0}

\par Let $X$ be a $2n$-dimensional closed spin manifold with a contractible dense subset $U$. Assume a coordinate system on $U$, say $(x^0, \cdots,x^{2n-1})$, and a Lorentzian metric $\eta$ on $X$ which is diagonal for coordinate system $x^\mu$, $\eta=\emph{\emph{diag}}(-1,1,\cdots,1)$.\footnote{For theoretical reasons we may also consider the time $t=x^0$ to vary from $-\infty$ to $+\infty$ on $U$.} Consider $X$ as a product manifold $X=Y \times Z$, where $Y$ is a closed $m$-dimensional manifold and $Z$ is a $2n'$-dimensional torus with radius $R$. We also suppose that $\{ x^\mu\}_{\mu=0}^{2n-1}$ splits $X$, that is $x^\mu$ belongs to $Y$ for $0 \leq \mu \leq m-1$, and $x^i$ is defined on the noncommutative manifold $Z$ for $m \leq i \leq 2n-1$. 

\par Actually, based on the above assumptions we redefine the subspace sweeped by the noncommutative coordinates $x^i$s, $m \leq i \leq 2n-1$, as a separate manifold $Z$. Obviously, the rest part of the spacetime manifold $X$, the so called $Y$, would be a commutative manifold, and hereby; $X=Y\times Z$. For instance, for the four-dimensional Moyal noncommutative spacetime with noncommutative coordinates $x^2$ and $x^3$, we obtain; $Y=\mathbb{R}^2$, spanned by $x^0$ and $x^1$, and $Z=\mathbb{R}^2_{\theta}$, including the noncommutative coordinates $x^2$ and $x^3$, wherein
$$\theta=\left( \theta^{ij} \right) \propto \left( {\begin{array}{*{20}{c}}
   0 & 1  \\
   { - 1} & 0  \\
\end{array}} \right),$$
\noindent for $i,j=2,3$, is the Moyal noncommutativity matrix. Also, to avoid the UV/IR mixing problem, we follow the technical trick of compactifying the noncommutative coordinates $x^2$ and $x^3$ \cite{ardalan, varshovi0, varshovi-consist}, hence the noncommutative part $\mathbb{R}^2_{\theta}$ is replaced by $Z=\mathbb{T}^2_{\theta}$, with $[x^2,x^3]=i \theta^{23}$. Moreover, for topological reasons, as it is convention in the literature, the extremely distant regions of the spacetime manifold $X$ must be compactified to a single point, the infinity. Therefore, we may assume $Y=S^2$ for the commutative part of spacetime. Altogether, the four-dimensional Moyal noncommutative spacetime is considered as $X=S^2 \times \mathbb{T}^2_\theta$ in our formalism. However, the formulations which we provide here can be easily extended to other dimensions of spacetime and to general translation-invariant noncommutative star products.\footnote{Moreover, as a special case, our algebraic framework could be generalized upon the viewpoints of the superstring theory introduced in \cite{witten}. For this objective one can also consider $Z=\mathbb{T}^6_{\star}$ as a six-dimensional noncommutative torus representing the extra dimensions (assumed at the limit of $R \to 0$ for the phenomenological reasons), where the background $B_{\mu \nu}$ field is living on. Then, $Y$ is considered as the ordinary four-dimensional observable spacetime. All in all, the formalism of which we are introducing here is flexible enough to contain various models of noncommutative structures.}

\par Set a translation-invariant star product $\star$ on $C^{\infty}(Z)$ via the cocycle $\alpha$ as \cite{lizzi, galluccio};
\begin{equation} \label {star}
(f\star g)(x):=\sum_{p,q \in \mathbb{Z}^{2n'}}~\tilde f(p) ~\tilde g(q)~e^{\alpha(p+q,p)}~e^{i(p+q).x/R}~,
\end{equation}

\noindent for
$$\tilde f(p)=\frac{{1}}{{  (2\pi R)^{2n'}  }}\int_{x \in Z}~f(x)~e_{-p}(x)~,~~~~~f(x)=\sum_{p\in \mathbb{Z}^{2n'}} \tilde f (p)~e_p(x)~,$$

\noindent the Fourier transformation on $Z$ with Fourier basis $e_p(x)=e^{ip.x/R}$, $p\in \mathbb{Z}^{2n'}$. Here, the integration is taken place with the Riemannian volume form $d^{2n'}x$. As a consequence of the Hodge theorem for the second group of $\alpha^*$-cohomology \cite {varshovi1, varshovi2} the 2-cocycle $\alpha$ is simply decomposed as $\alpha=\alpha_M+\partial \beta$, wherein $\alpha_M(p,q)=ip\wedge q:=-ip_iq_j\theta^{ij}/R^2$ is a Moyal 2-cocycle for the invertible noncommutativity matrix $\theta=(\theta^{ij})$, $m\leq i,j \leq 2n-1$, and $\beta \in C^{\infty}(Z)$ is a 1-cocycle ($\overline{\beta}(p)=\beta(-p)$ and $\beta(0)=0$) with coboundary
\begin{equation} \label {beta}
\partial \beta(p,q)=\beta (p-q)-\beta(p)+\beta(q)~.
\end{equation}

\noindent Therefore, if $\star_M$ is the Moyal product due to the 2-cocycle $\alpha_M$ we see;
\begin{equation} \label {1}
f' \star_M g' =(f \star g)'~,
\end{equation}

\noindent where
\begin{equation} \label {'}
f'(x)=\sum_{p\in \mathbb{Z}^{2n'}} \tilde f(p)~e^{\beta(p)}~e_p(x)~.
\end{equation}

\noindent Let $H$ be a separable Hilbert space with basis $\{ \left| p \right\rangle \}_{p\in \mathbb{Z}^{2n'}}$ and define $\pi: C^{\infty}(Z) \to \mathcal{L}(H)$ with
\begin{equation} \label {pi}
\pi(f)=\sum_{p \in \mathbb{Z}^{2n'}} \tilde f(p) ~ \pi_p~,~~~~~f\in C^{\infty}(Z)~,
\end{equation}

\noindent for $(\pi_p)_{r,s}=e^{ir\wedge s}~\delta_{p,r-s}$, with $r,s \in \mathbb{Z}^{2n'}$.\footnote{If we do not compactify $\mathbb{R}^{2n'}_{\star}$ to $Z=\mathbb{T}^{2n'}_{\star}$ in the noncommutative spacetime $X$, the Hilbert space $H$ in the formulations we provide here would no longer be separable. Then, its basis $\{ \left| p \right\rangle \}$ could be labeled with continuous $p \in \mathbb{R}^{2n'}$.} Obviously, $\pi$ is a $\mathbb{C}$-linear map and it can be seen that it provides a representation of noncommutative torus $C^{\infty}_{\star_M}(Z)=\mathbb{T}^{2n'}_{\theta}$ on $H$. That is;
\begin{equation} \label {rep}
\pi(f).\pi(g)=\pi(f \star_M g)~, ~~~~~f,g \in C^{\infty}(Z)~,
\end{equation}

\noindent where $.$ on the left hand side is the product of operators. Combining (\ref {rep}) with (\ref {'}) leads to a well-defined representation of $\star$ within a matrix base formulation. Denote $\pi(f')$ with $\hat f$, hence;
\begin{equation} \label {main}
\hat f.\hat g=\widehat {f\star g}~,~~~~~f,g \in C^{\infty}(Z)~.
\end{equation}

\noindent The representation map $\zeta: f \mapsto \hat f$ can be simply extended to functions $f  \in C^{\infty}(X)$. Let us show the algebra generated by $\hat f$, $f \in C^{\infty}(X)$, with $\frak{X}_0$. It is easily seen that $1=\hat 1$ and $\hat f^{\dag}=\widehat {\overline {f}}$. Therefore, $\frak{X}_0$ is in fact a unital $*$-algebra with involution $\hat f^*=\hat f^{\dag}$. Actually, since $1\star f =f$ we see that $C^{\infty}(X)=C^{\infty}_\star(X)$ as sets. This together with (\ref {main}) shows that $\zeta:C_\star^{\infty}(X) \to \frak{X}_0$ is an isomorphism of unital $*$-algebras.

\par The invertibility of $\zeta$ leads us to a natural definition for noncommutative polynomials. Assume that $P(x_1,\cdots,x_k)$ is a polynomial of probably noncommutative variables $x_i$s, then the noncommutative polynomial of $P$ for translation-invariant noncommutative star product $\star$, shown by $P_\star$, is defined as
\begin{equation} \label {polynomial}
P_\star(f_1,\cdots,f_k)=\zeta^{-1}\left( P(\hat f_1,\cdots,\hat f_k) \right)~,~~~~~f_1,\cdots,f_k \in C^{\infty}(X)~.
\end{equation}

\par The application of the representation map $\zeta$ is easily generalized to vector or matrix valued functions over $X$. Particularly, if $v=(v_i)$ is a vector valued function on $X$, such as a smooth section of a (trivial) vector bundle, it is mapped to $\hat v=(\hat v_i)$ by $\zeta$. Similarly, a matrix valued function, say $g=(g_{ij})$, is represented as $\hat g=(\hat g_{ij})$ via $\zeta$. Moreover, if $\{t^a\}$ is a set of definite matrices, such as a basis for a given Lie algebra, say $\frak{g}$, and $f=f^at^a$, $f^a \in C^{\infty}(X)$, is a smooth function with values in $\frak{g}$, then we would have; $\hat f =\hat f^at^a$.

\par The domain of the representation map $\zeta$ also can be extended to differential forms on $X$.\footnote{Actually, the definition of $\zeta$ can be simply generalized to any smooth tensor field on $X$ in the same way.} More precisely, if $\omega=f_{\mu_1 \cdots \mu_k}~dx^{\mu_1}\cdots dx^{\mu_k}$, for smooth functions $f_{\mu_1 \cdots \mu_k}$s, totally anti-symmetric in indices $\mu_i$, is mapped to $\hat \omega=\hat {f}_{\mu_1 \cdots \mu_k}~dx^{\mu_1}\cdots dx^{\mu_k}$ through with $\zeta$. These matrix valued differential forms can be multiplied with the common sense due to the operator multiplication and the wedge product of differential forms. We show this algebra with $\frak{X}$, which is in fact a graded algebra as $\frak{X}=\oplus_{k=0}^{2n}\frak{X}_{k}$, wherein $\frak{X}_k$, $0 \leq k \leq 2n$, is the image of the differential $k$-forms on $X$ via $\zeta$, i.e. the space of operator valued $k$-forms on $Y$ with extra dimensions on $Z$. In the following we will show that, it is in principle a differential graded algebra.

\par The elements of $\frak{X}_0$ can be partially differentiated with respect to $x^{\mu}$, $0 \leq \mu \leq m-1$, on $Y$ with the usual way. However, although these elements have missed their dependence to the points of $Z$, but the partial differentiation with respect to $x^i$s, $m\leq i \leq 2n-1$, is generalized to such elements. We define;
\begin{equation} \label {partial}
\partial_i\hat f=\widehat{\partial_i f}~.
\end{equation}

\noindent This lets us to obtain an exterior derivative operator on $\frak{X}$ as
\begin{equation} \label {d}
d\hat \omega=\sum_{\mu =0}^{2n-1} \partial_{\mu}\hat f_{\mu_1 \cdots \mu_k}~dx^{\mu}dx^{\mu_1}\cdots dx^{\mu_k}~,
\end{equation}

\noindent where $\omega=f_{\mu_1 \cdots \mu_k}~dx^{\mu_1}\cdots dx^{\mu_k}$. Obviously we have;
$$d^2=0~,~~~~~d(\hat \omega_1.\hat \omega_2)=d\hat \omega_1.\hat \omega_2+(-1)^{|\hat \omega_1|}\hat \omega_1.d\hat \omega_2~,$$
\noindent where $|\hat \omega|$ is the degree of $\hat \omega$ in $\frak{X}$. This shows that $(\frak{X},d)$ is a differential graded algebra over $\frak{X}_0$. Let us refer to the cohomology theory of $(\frak{X},d)$ as the $\star$-cohomology and denote its groups with $H_\star^k(X,\mathbb{C})$, $k\geq 0$. Since $\zeta \circ d_X =d \circ \zeta$, for $d_X$ the exterior derivative on $X$, the following diagram commutes;
\begin{equation} \label {diagram}
\begin{array}{*{20}{c}}
 \cdots & \to &   \Omega_{k-1}(X) & \to &   \Omega_{k}(X) & \to  &  \Omega_{k+1}(X) & \to & \cdots  \\
 ~ & ~ & \zeta \downarrow & {~} & \zeta \downarrow  & {~} &  \zeta \downarrow & ~ & ~   \\
 \cdots & \to &  \frak{X}_{k-1} & \to  &  \frak{X}_{k} & \to  & \frak{X}_{k+1} & \to & \cdots  \\
\end{array}~,
\end{equation}

\noindent wherein the upper and the lower arrows represent $d_X$ and $d$ respectively, while $\Omega_*(X)$ is the ordinary de Rham complex. Hence, $\zeta$ induces a homomorphism as $\zeta_*:H_{dR}^k(X,\mathbb{C}) \to H_\star^k(X,\mathbb{C})$ for each $0\leq k \leq 2n$. However, since $\zeta:C_\star^{\infty}(X) \to \frak{X}_0$ is an isomorphism then the column maps of (\ref {diagram}) are all bijective. Therefore, we have already established the following statement. \\

\par \textbf{Proposition 1;} \emph{Assume that $\star$ is a translation-invariant noncommutative star product on $X$. Then, the isomorphism $\zeta:C^{\infty}_\star(X) \to \frak{X}_0$ induces an isomorphism $\zeta_*:H_{dR}^k(X,\mathbb{C}) \to H_\star^k(X,\mathbb{C})$ for each $k\geq 0$.}\\

\par Before closing this section let us prove a useful lemma.\\

\par \textbf{Lemma 1;} \emph{Consider the transformation (\ref {'}) for 1-cocycle $\beta$. Hence if there exists a fixed $C\in \mathbb{C}$ so that for any set of functions $f_1,\cdots,f_k \in C^{\infty}(X)$ the following equality holds;}
\begin{equation} \label {lemma1}
\int_X f'_1 \cdots  f'_k=C\int_X f_1 \cdots f_k~,
\end{equation}

\noindent \emph{then, $C$ must be equal to the unity and $\beta$ has to be linear.}\\

\par \textbf{Proof;} Writing down the equality of (\ref {lemma1}) on the Fourier space the integral on the right hand side receives a coefficient of $e^{\sum_i \beta(p_i)}\delta_{\sum_{i}p_i,0}$ which must be equal to $C$ for any set of Fourier modes $\{p_i\}_{i=1}^k$ including $p_i=0$, $1\leq i \leq k$. That is, $C=e^{\sum_i \beta(0)}=e^{\beta(p)+\beta(-p)}=e^{\beta(p)+\beta(q) + \beta(-p-q)}$. Therefore, since $\beta(0)=0$, then $C$ must be equal to the unity and thus we obtain $\beta(p)=-\beta(-p)$ and $\beta(-p-q)=-\beta(p)-\beta(q)$ for any $p,q \in \mathbb{Z}^{2n'}$. Hence $\beta:\mathbb{Z}^{2n'}\to \mathbb{C}$ is a linear function. \textbf{Q.E.D}


\par
\section{$\star$-Cohomology and the Third Type Chern Character}
\setcounter{equation}{0}

\par However, we are mostly interested in integral cohomology group $H_\star^{2n}(X,\mathbb{Z})$. To define it we need some integration structure on $\frak{X}$. It is simply defined as;
\begin{equation} \label {integral}
\hat{\int} \hat \omega =
\left\{ {\begin{array}{*{20}{c}}
   \int_Xf~,~~~~~~~~~~\hat \omega =\hat f ~d^{2n}x \in \frak{X}_{2n}~~~ \\
   0 ~,~~~~~~~~~~~~~~~~~~~~~\emph{\emph{otherwise}}~~~~~~~~ \\
\end{array}} \right.
~,
\end{equation}

\noindent where the integral on $X$ is taken for the Riemann volume form $d^{2n} x=dx^0 \cdots dx^{2n-1}$. We see
$$\hat {\int} d\hat \omega=0~,~~~~~\hat \int \hat{\omega}_1.\hat{\omega}_2=(-1)^{n_1n_2}\hat \int \hat {\omega}_2.\hat{\omega}_1~, ~~~~~\hat{\omega}_i \in \frak{X}_{n_i}(X)~,~i=1,2~,$$
\noindent which shows $\hat \int$ is a closed graded trace over $(\frak{X},d)$ and thus represents an element of cyclic cohomology group $HC^{2n}(\frak{X}_0)$ \cite {connes-book}. It is obvious that if $\hat \omega \in \frak{X}_{2n}$ is exact then $\hat \int \hat \omega=0$ due to closedness of $X$. Thus, in fact, the integral structure (\ref {integral}) is invariant on the cohomology classes of $H_\star^{2n}(X,\mathbb{C})$. The space of the cohomology classes in $H_\star^{2n}(X,\mathbb{C})$ with integer integrals due to (\ref{integral}) are then referred to with $H_\star^{2n}(X,\mathbb{Z})$, the so called integral group of the $\star$-cohomology. But since $\hat \int \circ \zeta =\int_X$ and $\zeta:\Omega_{2n}(X) \to \frak{X}_{2n}$ is bijective, we readily find; \\

\par \textbf{Proposition 2;}  \emph{Assume that $\star$ is a translation-invariant noncommutative star product on $X$. Then, the isomorphism $\zeta:C^{\infty}_\star(X) \to \frak{X}_0$ induces an isomorphism on $2n^{th}$ integral cohomology groups $\zeta_*:H_{dR}^{2n}(X,\mathbb{Z}) \to H_\star^{2n}(X,\mathbb{Z})$. In principal; $H_\star^{2n}(X,\mathbb{Z})=\mathbb{Z}$.}\\

\par It is well know that if $E \to X$ is a vector bundle over $X$, then its highest Chern character $\text{ch}_{n}(E)$ provides a definite class in $H^{2n}_{dR}(X,\mathbb{Z})$. Moreover, it is seen that $H^{2n}_{dR}(X,\mathbb{Z})$ is generated with such Chern characters of the vector bundles over $X$. Therefore, we readily have the following corollary;\\

\par \textbf{Corollary 1;} \emph{The integral cohomology group $H_\star^{2n}(X,\mathbb{Z})$ is generated by $\hat{\emph{{ch}}}_{n}(E)=\zeta(\emph{{ch}}_{n}(E))$ for all vector bundles $E \to X$.} \\

\par Let us refer to $\hat{\emph{\emph{ch}}}_{2n}(E)$ as the first type Chern character and show it by $\hat{\emph{\emph{ch}}}^{(1)}_{2n}(E)$ to emphasize on the term of \emph{first}. But, however, applying the noncommutativity of $\star$ we can define more types of Chern characters in integral cohomology group $H_\star^{2n}(X,\mathbb{Z})$. To define them we need some more simplifying assumptions. Since $U$ is contractible then any vector bundle $\mathbb{C}^k \to E \to X$ has a trivialization over it. Thus, any covariant derivative on it, say $\nabla$, is given as $d_X+A$, for $A$ a differential one-form over $U $ with values in $\mathbb{M}_{k}(\mathbb{C})$. Its curvature is then a $\mathbb{M}_{k}(\mathbb{C})$-valued two-form as; $\nabla^2=d_X A + A^2=\frac{{1}}{{2}} \left( \partial_{ [ \mu } A_{\nu ]} + [A_{\mu},A_{\nu}] \right)~dx^{\mu}dx^{\nu}=\frac{{1}}{{2}}F_{\mu \nu}~ dx^{\mu}dx^{\nu}$. Then, the second type Chern character $\hat{\emph{\emph{ch}}}^{(2)}_n(E)$ is defined with: $\frac{{1}}{{n!}}\left( \frac{{i}}{{4 \pi }} \right)^n Tr\{ (\hat {F}_{\mu\nu}~dx^{\mu}dx^\nu)^{n}\}$.\footnote{From now on, $Tr$ is the trace on the components of $\mathbb{C}^k$ in $\mathbb{C}^k \to E\to X$.} In fact, $\hat{\emph{\emph{ch}}}^{(2)}_n(E)$ is the Chern character for the mapped curvature $\frac{{1}}{{2}}\hat {F}_{\mu\nu}~dx^{\mu}dx^\nu$. For instance;
$$\hat{\emph{\emph{ch}}}_1^{(2)}(E)=\frac{{i}}{{4\pi}}Tr\{\hat {F}_{\mu\nu} \}~\epsilon^{\mu \nu}~d^2x~, ~~~~~~~~~~\hat{\emph{\emph{ch}}}_2^{(2)}(E)=-\frac{{1}}{{32\pi^2}}Tr\{\hat F_{\mu\nu}\hat F_{\sigma\lambda}\}~\epsilon^{\mu \nu \sigma \lambda}~d^4 x~,$$
\noindent for $\epsilon^{\mu \nu}$ and $\epsilon^{\mu\nu\sigma\lambda}$ the Levi-Civita tensors with $\epsilon^{01}=1$ and $\epsilon^{0123}=1$. It is obvious that $\hat \int \hat{\emph{\emph{ch}}}_1(E)$ is an integer, since it has no involved $\star$. Thus, we are in fact integrating the Chern character in a normal way as we apply $\hat \int$. But, actually, this coincidence is exceptional only for 2 dimensions unless we restric ourselves to the Moyal products. Here we find the following theorem. \\

\par \textbf{Theorem 1;} \emph{Suppose that $\star$ is a translation-invariant noncommutative star product on $X$, $E$ is a vector bundle over it, and $\nabla$ is a connection on $E$. If $n\geq 2$ then, $\star$ is a Moyal product if and only if $\emph{ch}^{(2)}_n(E)$ defines a cohomology class in $H_\star^{2n}(X,\mathbb{C})$ independent of the connection $\nabla$. Hence after;}
\par \textbf{a)} \emph{The cohomology class of $\hat{\emph{ch}}^{(2)}_n(E)$ belongs to $H_\star^{2n}(X,\mathbb{Z})$.}
\par \textbf{b)} \emph{$\hat{\emph{ch}}^{(2)}_n(E)$ and $\hat{\emph{ch}}_n^{(1)}(E)$ are cohomologous. That is;}
\begin{equation} \label {prop1-0}
 \hat {\int} \hat{\emph{\emph{ch}}}^{(2)}_n(E)=\hat \int \hat {\emph{\emph{ch}}}^{(1)}_n(E)=\int_X \emph{\emph{ch}}_n(E) ~\in \mathbb{Z}~.\\
\end{equation}
~
\par \textbf{Proof;} The "if" part has already been established in \cite {varshovi-first} where we studied the second type Chern character for the Moya star product. For the "only if" part assume that $\star$ is given for 2-cocycle $\alpha=\alpha_M+\partial \beta$ due to the Hodge decomposition in $\alpha^*$-cohomology. Thus, we should show that if the cohomology class of $\hat{\emph{\emph{ch}}}^{(2)}_n(E)$ for $\nabla=d_X+A$ is independent of the connection form $A$, then $\partial \beta=0$. To do this one should initially note that for any set of functions $f_1,\cdots,f_k \in C^{\infty}(X)$ we have;
\begin{equation} \label {prop1-1}
\int_X f_1\star \cdots \star f_k=\int_X (f_1\star \cdots \star f_k)'=\int_X f'_1\star_M \cdots\star_M f'_k~,
\end{equation}

\noindent where the first equality comes from $\beta(0)=0$ and the second one is a direct consequence of the iterating formula (\ref {1}). Therefore, we read form (\ref {prop1-1});
\begin{equation} \label {theorem1-1}
\hat \int \hat{\emph{\emph{ch}}}^{(2)}_n(E)=\left(\frac{{i}}{{4\pi}} \right)^{n}\int_X Tr\{ F'_{\mu_1\nu_1}\star_M \cdots \star_M F'_{\mu_n \nu_n} \} \epsilon^{\mu_1\nu_1 \cdots \mu_n \nu_n}~.
\end{equation}

\noindent In \cite {varshovi-first} we proved that the Moyal product can be replaced by the ordinary product in (\ref {theorem1-1}).\footnote{The proof of this claim is actually based on a proposition which states that the integral of a symmetric production of a set of smooth functions multiplied with the Myal product $\star_M$ is independent of the values of the noncommutativity matrix entries $\theta^{ij}$.} Thus;
\begin{equation} \label {theorem1-2}
\hat \int \hat{\emph{\emph{ch}}}^{(2)}_n(E)=\left(\frac{{i}}{{4\pi}} \right)^{n}\int_X Tr\{ F'_{\mu_1\nu_1} \cdots  F'_{\mu_n \nu_n} \} \epsilon^{\mu_1\nu_1 \cdots \mu_n \nu_n}~.
\end{equation}

\noindent If the result of (\ref {theorem1-2}) is independent of the connection form $A$, then it must be a constant, just similar to the topological invariant $\int \emph{\emph{ch}}_n(E)$. That is, there exists a fixed $C\in \mathbb{C}$ so that;
\begin{equation} \label {theorem1-2newtar}
\int_X Tr\{ F'_{\mu_1\nu_1} \cdots  F'_{\mu_n \nu_n} \} \epsilon^{\mu_1\nu_1 \cdots \mu_n \nu_n}=C\int_X Tr\{ F_{\mu_1\nu_1} \cdots  F_{\mu_n \nu_n} \} \epsilon^{\mu_1\nu_1 \cdots \mu_n \nu_n}~.
\end{equation}

\noindent Now by \textbf{Lemma 1} we find $C=1$ and obtain $\beta$ as a linear function. Consequently, $\partial \beta$ vanishes and indeed $\alpha=\alpha_M$. This proves (\ref {prop1-0}). The rest of the proof is clear due to the Chern-Weil theory. Actually, since $\hat {\int} \hat{\emph{\emph{ch}}}^{(2)}_n(E)$ is an integer, then by definition $\hat{\emph{\emph{ch}}}^{(2)}_n(E)$ defines a definite class of $H_\star^{2n}(X,\mathbb{Z})$. Moreover, due to the equality $\hat {\int} \hat{\emph{\emph{ch}}}^{(2)}_n(E)=\hat \int \hat {\emph{\emph{ch}}}^{(1)}_n(E)$ we conclude that $\hat{\emph{\emph{ch}}}^{(2)}_n(E)-\hat {\emph{\emph{ch}}}^{(1)}_n(E)$ is an exact form, hence $\hat{\emph{\emph{ch}}}^{(2)}_n(E)$ and $\hat {\emph{\emph{ch}}}^{(1)}_n(E)$ must belong to the same class. \textbf{Q.E.D}\\

\par We remember that the Moyal and the Wick-Voros star products are cohomologous 2-cocycles in the sense of $\alpha^*$-cohomology \cite{varshovi1, varshovi2}. On the other hand, due to the above theorem the second type Chern characters for the Moyal and the Wick-Voros star products belong generally to different cohomology classes in $H^{2n}_\star(X,\mathbb{C})$. This is while it has already been shown that the whole quantum features (including the topological structures) of two versions of a quantum field theory described with two different $\alpha^*$-cohomologous star products coincide precisely \cite{varshovi2}. Therefore, the second type Chern character is not in fact a topological invariant of the theory except for the Moyal product. 

\par However, there exists a third type of Chern character for translation-invariant noncommutative structures on $X$ which we refer to as the \emph{gauge equivariant} Chern character. To define this one should consider the image of the connection on the vector bundle $E \to X$ through $\zeta$. This is in fact given as $\hat {\nabla}=d+\hat A$, for $\nabla = d_X +A$. We define the image of the vector bundle $E \to X$ via $\zeta$ as the vector space of
\begin{equation} \label {ehat}
\hat E:=\{\hat \sigma=\zeta(\sigma)|~\sigma~\emph{\emph{a~smooth~global~section~of}}~E\}~,
\end{equation}

\noindent and then define $\hat \nabla$ as a covariant derivative with $\hat \nabla \hat \sigma=d\hat \sigma + \hat A \triangleright \hat \sigma$, wherein the symbol $\triangleright$ determines the type of the action of $\hat A$ on $\hat \sigma$. In the next section we will argue more about the role of this action in translation-invariant noncommutative Yang-Mills theories. But,  the curvature $\hat \nabla^2$ is actually defined regardless of the type of the action $\triangleright$ and is given by $\hat {\nabla}^2=d\hat A + \hat{A}^2$ as an element of $\frak{X}_2$.

\par The \emph{third type Chern character} is defined as $\hat{\emph{\emph{ch}}}^{(3)}_*(E)=Tr\{ \exp (\frac{{i}}{{2\pi}} \hat {\nabla}^2 ) \}$. It is with more interest in translation-invariant noncommutative gauge theories because of its gauge covariance (or gauge invariance under integration). Moreover, it is intimately correlated to the Connes-Chern character in the cyclic (co)homology in noncommutative geometry \cite {varshovi-forth}. We are mostly interested in the highest term of the third type Chern character $\hat{\emph{\emph{ch}}}^{(3)}_n(E)$ due to its topological properties (as we will establish in the next theorem). Let us show it with the simpler notation of $\check{\emph{\emph{ch}}}_n(E)$. Therefore, for the $D$-dimensional vector bundle $\mathbb{C}^D \to E \to X$ we have:
$$\check{\emph{\emph{ch}}}_1(E)=\frac{{i}}{{4\pi}}Tr\{ \hat {F}_{\star \mu\nu }\}~\epsilon^{\mu \nu}~dx^0dx^1~,$$
\noindent on two-dimensional manifold $X$, and
$$\check{\emph{\emph{ch}}}_2(E)=-\frac{{1}}{{32\pi^2}} Tr\{ \hat{F}_{\star \mu\nu}\hat{F}_{\star \sigma \lambda} \}~\epsilon^{\mu \nu \sigma \lambda}~dx^0dx^1dx^2dx^3~,$$

\noindent on four-dimensional spacetime $X$, wherein
\begin{equation} \label {Rstar}
F_{\star \mu \nu}=F^a_{\star\mu \nu}~ t^a=\partial_{[\mu}A^a_{\nu ]}~t^a+\frac{{1}}{{2}}  \{A_{\mu}^a , A_{\nu}^b\}_\star  ~f^{ab}_c~t^c+\frac{{1}}{{2}}[A^a_{\mu},A^b_{\nu}]_{\star}~\{t^a,t^b\}~.
\end{equation}

\noindent for $\{t^a\}$ a basis of $\frak{gl}(D,\mathbb{C})$ with structure constant $f^{ab}_c$. The next theorem shows the significance of $\check{\emph{\emph{ch}}}_n(E)$ in $H_\star^{2n}(X,\mathbb{Z})$.\\

\par \textbf{Theorem 2;} \emph{Consider a vector bundle, say $E \to X$, together with a general translation-invariant noncommutative star product $\star$ on $X$. Then, the following properties hold;}
\par \textbf{a)} \emph{$\check{\emph{ch}}_n(E)$ defines an integral cohomology class in $H_{\star}^{2n}(X,\mathbb{Z})$.}
\par \textbf{b)} \emph{The cohomology class of $\check{\emph{ch}}_n(E)$ is independent of the connection $\nabla$.}
\par \textbf{c)} \emph{Moreover, $\check{\emph{ch}}_n(E)$ and $\hat{\emph{ch}}^{(1)}_n(E)$ are cohomologous, i.e.;}
\begin{equation} \label {theorem5}
\hat {\int} \check{\emph{\emph{ch}}}_n(E)= \hat {\int} \hat{\emph{\emph{ch}}}^{(1)}_n(E)=\int_X \emph{\emph{ch}}_n(E) ~ \in \mathbb{Z}~.
\end{equation}\\

\par \textbf{Proof;} Let us show the third type Chern character for $\hat{\nabla}=d+\hat A$ with notation $\check{\emph{\emph{ch}}}_n(E,A)$. Then, due to (\ref {1}) we obtain;
\begin{equation} \label {theorem2-1}
\hat \int \check{\emph{\emph{ch}}}_n(E,A)=\int_X \check{\emph{\emph{ch}}}_{n\star}(E,A)=\int \check{\emph{\emph{ch}}}_{n\star_M}(E,A')~,
\end{equation}

\noindent for noncommutative polynomials $\check{\emph{\emph{ch}}}_{n\star}(E,A)$ and $\check{\emph{\emph{ch}}}_{n\star_M}(E,A')$ due to (\ref {polynomial}). The Moyal product $\star_M$ can also be replace by the ordinary product under integration of $\check{\emph{\emph{ch}}}_{n\star_M}(E,A')$ in (\ref {theorem2-1}) \cite{varshovi-first}. Thus;
\begin{equation} \label {theorem2-2}
\int_X \check{\emph{\emph{ch}}}_{n\star_M}(E,A')=\int_X {\emph{\emph{ch}}}_{n}(E,A')=\int_X {\emph{\emph{ch}}}_{n}(E,A)=\int_X {\emph{\emph{ch}}}_{n}(E)~,
\end{equation}

\noindent where the second and the third equalities are immediate consequences of the Chern-Weil theory. More precisely, according to the Chern-Weil theory the topological index $\int_X {\emph{\emph{ch}}}_{n}(E,A)$ is independent of the connection form $A$, hence the second equality (and the third one) of (\ref{theorem2-2}) follows consequently. This finishes the theorem. \textbf{Q.E.D}


\par
\section{Third Type Chern Character and Noncommutative Yang-Mills}
\setcounter{equation}{0}

\par The last theorem of the previous section has a significant role in studying the Yang-Mills theories with genral translation-invariant noncommutative star products $\star$. Such theories are defined by $\hat \nabla =d+\hat A$ via\footnote{For the case of noncommutative QED one should use an extra overal factor of $\frac{{1}}{{2}}$ \cite {bertlman}.}
\begin{equation} \label {ymaction}
\mathcal{S}_{Y-M}=\hat \int \hat L_{Y-M}:=\hat \int Tr\{\hat {\nabla}^2 * (\hat {\nabla}^2)\}=-\frac{{1}}{{2}}\int_X~Tr\{F_{\star\mu\nu}F_\star^{\mu\nu}\}=-\frac{{1}}{{4}}\int_X~F^a_{\star\mu\nu}F_\star^{a\mu\nu}
\end{equation}

\noindent for the \emph{Yang-Mills $\zeta$-Lagrangian} $\hat L_{Y-M}$, the Hodge star operator $*$, and $\hat \nabla ^2=-\frac{{i}}{{2}}\hat F_\star=-\frac{{i}}{{2}} \hat {F}_{\star \mu \nu}~dx^\mu dx^\nu$ where\footnote{To have a more concrete formulation, from now on we write the connection form $A_\mu$ and its curvature $F_{\mu\nu}$ with an overal factor of $-i$, i.e. $A_\mu \to -i A_\mu$ and $F_{\mu\nu} \to -i F_{\mu\nu}$.}
\begin{equation} \label {fstar}
F_{\star \mu \nu}=F^a_{\star\mu \nu}~ t^a=\partial_{[\mu}A^a_{\nu ]}~t^a+\frac{{1}}{{2}}  \{A_{\mu}^a , A_{\nu}^b\}_\star  ~f^{abc}~t^c+\frac{{i}}{{2}}c^{abc}[A^a_{\mu},A^b_{\nu}]_{\star}~,
\end{equation}

\noindent with $[t^a,t^b]=if^{abc}~t^c$ and $\{t^a,t^b\}=-c^{abc}~t^c$. Accordingly, $Tr$ in (\ref {ymaction}) is the trace over color indices of the internal gauge symmetry. Note that here we assume the gauge group $G$ is generated with anti-hermitian generators $-it^a$ in some appropriate representation so that the corresponding Lie algebra $\frak{g}$ is closed for the anti-commutators $\{t^a,t^b\}$. Hence, we may assume $G=U(N)=U(1)\times SU(N)$, for some $N$, in the fundamental representation, where $t^0=\frac{{1}}{{\sqrt{2N}}}\mathbb{I}$ and $t^a $s are the Hermitian generators of $\frak{su}(N)$, $1 \leq a \leq N^2-1$. Thus, we have; $Tr\{t^at^b\}=\frac{{1}}{{2}}\delta^{ab}$ for $0 \leq a,b \leq N^2-1$.

\par The Dirac action is also given for the \emph{Dirac $\zeta$-Lagrangian} $\hat L_{D}$ as
\begin{equation} \label {dirac}
\mathcal{S}_{D}=\hat \int \hat L_{D}:=i\hat \int \overline {\hat{\psi}} \gamma^\mu {\partial_\mu} \hat{\psi}~d^{2n}x=i\int_X\overline{\psi} \gamma^\mu \partial_\mu \psi~,
\end{equation}

\noindent with $\hat {\psi}$ the image of the spinor field $\psi$ via the representation map $\zeta$, and $\overline{\hat{\psi}}=\hat{\psi}^{\dag}\gamma^0$. The matter fields interact with the gauge fields in three different types:
\begin{equation} \label {matter}
\begin{array}{*{20}{c}}
    \bold{Type~A:~Ordinary~Action;} & ~~\mathcal{S}_{int}^A=\hat {\int} \overline {\hat{\psi}}\gamma_\mu t^a \hat A^a_\mu  \hat{\psi} ~d^{2n}x =\int_X \overline{\psi}\gamma_\mu t^a \star A^a_\mu \star \psi ~, \\
    \bold{Type~B:~Inverse~Action;}~~~ & ~~~~~~\mathcal{S}_{int}^B=-\hat {\int} \overline {\hat{\psi}}\gamma_\mu  t^a \hat{\psi} \hat A^a_\mu  ~d^{2n}x = - \int_X \overline{\psi}\gamma_\mu t^a \star \psi \star A^a_\mu ~, \\
   \bold{Type~C:~Adjoint~Action;} ~~ & ~~~~~~\mathcal{S}_{int}^C=\hat {\int} \overline {\hat{\psi}}\gamma_\mu  t^a [\hat A^a_\mu , \hat{\psi}]  ~d^{2n}x = \int_X \overline{\psi}\gamma_\mu t^a \star [ A^a_\mu , \psi]_\star ~, \\
\end{array}
\end{equation}

\noindent which we simply write as $\mathcal{S}_{int}^T=\hat \int \hat L^T_{int}$ with $T=A,B,C$ for the \emph{interactin $\zeta$-Lagrangian} $\hat L^T_{int}$. Such theories have been studied extensively in the literature.\footnote{see \cite{ardalan, ardalan2, witten, varshovi-consist, varshovi0} and the references therein.} It is seen that the type A and the type B theories lead to the standard Yang-Mills theories in the commutative case of four-dimensional spacetime. However, although including rather complicated interaction terms, the type C theory leads to a pure gauge theory for the commutative fields with no coupling to the fermions.

\par Hence:
$$\mathcal{S}^A_{int}=\int_X A_\mu^a~ j_A^{a\mu}~,~~\mathcal{S}^B_{int}=-\int_X A_\mu^a~ j_B^{a\mu}~,~~\mathcal{S}^C_{int}=\mathcal{S}^A_{int}+\mathcal{S}^B_{int}=\int_X A_\mu^a~ j_C^{a\mu}~.$$
\noindent The classical conservation laws for currents $j^{a\mu}_{T}$, $T=A,B,C$, are
\begin{equation} \label {conservation1}
\begin{array}{*{20}{c}}
    \bold{Type~A:}~~~~~\partial_\mu j_A^{a\mu}+\frac{{1}}{{2}}f^{abc}\{A^b_\mu,j_A^{c\mu}\}_\star+\frac{{i}}{{2}}c^{abc}[A^b_\mu,j_A^{c\mu}]_\star =0 ~,~~~~~~~~~~~~~~~~~~~~~~~~~~~~~~~~~~~~~~~~~ \\
    \bold{Type~B:}~~~~~\partial_\mu j_B^{a\mu}-\frac{{1}}{{2}}f^{abc}\{A^b_\mu,j_B^{c\mu}\}_\star+\frac{{i}}{{2}}c^{abc}[A^b_\mu,j_B^{c\mu}]_\star =0  ~,~~~~~~~~~~~~~~~~~~~~~~~~~~~~~~~~~~~~~~~~~ \\
   \bold{Type~C:}~~~~~\partial_\mu j_C^{a\mu}+\frac{{1}}{{2}}f^{abc}\{A^b_\mu,j_A^{c\mu}+j_B^{c\mu}\}_\star +f^{abc}\gamma^\mu_{ji}~t^c_{\beta \alpha} \left(\psi_{i,\alpha}\star A^b_\mu \star \overline{\psi}_{j,\beta}+ \overline{\psi}_{j,\beta}\star A^b_\mu \star \psi_{i,\alpha} \right)\\
   +\frac{{i}}{{2}}c^{abc}[A^b_\mu,j_C^{c\mu}]_\star =0   ~.~~~~~~~~~~~~~~~~~~~~~~~~~~~~~~~~~~~~~~~~~~~~~~~~~~~~~~~~~ \\
\end{array}
\end{equation}

\noindent The axial currents $j^{a\mu}_{T5}$, $T=A,B,C$, also satisfy similar classical equations as (\ref {conservation1}) via replacing $\gamma^\mu \to \gamma^\mu \gamma_5$. Hence, the chiral singlet currents which are defined by replacing $t^a$ and $\psi$ respectively with $\mathbb{I}$ and the chiral fermion $\psi_H$, for chirality $H=L,R$, are conserved classically as;
\begin{equation} \label {conservation}
\partial_\mu j_{T,H}^{\mu}+i[A^b_\mu,j_{T,H}^{b\mu}]_\star =0~,~~~~~T=A,B,C~,
\end{equation}

\noindent wherein we have a summation over $0\leq b \leq N^2-1$.

\par Actually, all three actions of (\ref {matter}) are consistent representations of matrix valued gauge fields on the space of matrix valued spinors. But, as it is seen from (\ref {matter}) these types of representations give rise to three different forms of connections $\hat {\nabla}^A$, $\hat {\nabla}^B$, and $\hat {\nabla}^C$, on $\hat E$ as the matrix formulation of $E=\mathbb{C}^k \otimes S(X) \to X$. Note that here $\mathbb{C}^k$ is the representing space of the gauge group $G$ and $S(X) \to X$ is the spin bundle.

\par However, as it was mentioned in previous section, despite of different forms of the connections all the three representations of (\ref {matter}) lead to the same curvature $-\frac{{i}}{{2}}\hat {F}_{\star \mu \nu}~dx^\mu dx^\nu$, with $F_{\star \mu \nu}$ given as (\ref{fstar}). Consequently, they have the same third type Chern characters. Hence, the third type Chern characters are stable with respect to the connections, the gauge transformations,\footnote{We emphasize that $\check{\emph{\emph{ch}}}_n(E)$ is gauge covariant and also gauge invariant under integration.} the analytic form of translation-invariant noncommutative star products\footnote{Due to \textbf{Theorem 2} this is despite of the second type Chern character which defines topological structure of the corresponding theory only for the Moyal star product.} and the type of the actions. Essentially, it might be expected that they would play some significant physical roles in studying the theory of general translation-invariant noncommutative gauge fields. In the next theorem we prove that the anomaly of such theories are given in terms of $\check{\emph{\emph{ch}}}_{n\star}(E)$, where $\check{\emph{\emph{ch}}}_{n\star}(E)$ is the noncommutative polynomial due to (\ref {polynomial}).
\par However, before stating the next theorem let us have a quick review on noncommutative anomalies. Actually, if (\ref {conservation}) receives any anomalous contribution at quantum levels on the right hand side, say $\mathcal{A}_{T,H}$, then the chiral singlet charge $Q_{T,H}$ increases/decreases via equation
\begin{equation} \label {charge}
\Delta Q_{T,H}=Q_{T,H}(+\infty)-Q_{T,H}(-\infty)=\int_{-\infty}^{\infty}\frac{{d}}{{dt}}Q_{T,H}(t)=\int_X~\mathcal{A}_{T,H}
\end{equation}

\noindent form the extreme past to the extreme future. The physical consistency implies that such changes must be an integer times the unit charge of the theory, which consequently yields \cite {weinberg, bertlman, nakahara}:
\begin{equation} \label {consistency}
\mathcal{A}_{T,H} \in H_{dR}^{2n}(X,\mathbb{Z})~,~~~~~T=A,B,C~,~~~~~H=L,R~.
\end{equation}

\noindent The differential form $\mathcal{A}_{T,H}$ is commonly known as the chiral singlet or the chiral Abelian anomaly. The following theorem describes $\mathcal{A}_{T,H}$ in terms of the highest third type Chern character. \\

\par \textbf{Theorem 3;} \emph{Assume that $G=U(N)$, $N \ge 1$, and $ E \to X$ is a vector bundle with structure group $G$ in the fundamental representation. Moreover, consider a general translation-invariant noncommutative star product $\star$ on $X$. Then, due to the physical consistency (\ref {consistency}) we find;}
\par \textbf{a)} \emph{If $G=U(1)$, then the chiral anomaly of the corresponding translation-invariant noncommutative QED in the ordinary action (resp. the inverse action) is $ \mp \check{\emph{ch}}_{n\star}(E)$ (resp. $\pm \check{\emph{ch}}_{n\star}(E)$). Strictly speaking we have;}
\begin{equation} \label {theorem6a}
\partial_\mu j^\mu_{A,H}+i[A_\mu,j^\mu_{A,H}]_\star=\varepsilon_H\check{\emph{\emph{ch}}}_{n\star}(E)~,~~~~~(resp.~\partial_\mu j^\mu_{B,H}+i[A_\mu,j^\mu_{B,H}]_\star=-\varepsilon_H\check{\emph{\emph{ch}}}_{n\star}(E))~,
\end{equation}

\noindent \emph{for $\varepsilon_L=-1$ and $\varepsilon_R=+1$. Moreover, the translation-invariant noncommutative QED in the adjoint action is anomaly free.}
\par \textbf{b)} \emph{If $G$ is non-Abelian, the chiral Abelian anomaly of the corresponding translation-invariant noncommutative Yang-Mills theory in the ordinary action (resp. the inverse action) is $\mp \check{\emph{ch}}_{n\star}(E)$ (resp. $\pm \check{\emph{{ch}}}_{n\star}(E)$). That is,}
\begin{equation} \label {theorem6b}
\partial_\mu j^\mu_{A,H}+i[A^b_\mu,j^{b\mu}_{A,H}]_\star=\varepsilon_H \check{\emph{\emph{ch}}}_{n\star}(E)~,~~~~~(resp.~\partial_\mu j^\mu_{B,H}+i[A^b_\mu,j^{b\mu}_{B,H}]_\star=-\varepsilon_H\check{\emph{\emph{ch}}}_{n\star}(E))~.
\end{equation}

\noindent \emph{In addition, the translation-invariant noncommutative Yang-Mills theory in the adjoint action is anomaly free for the chiral singlet current.} \\

\par \textbf{Proof;} We prove \textbf{(b)}. The part \textbf{(a)} then follows consequently. Assume that the 2-cocycle of $\star$ is $\alpha$. We define a homotopy from the ordinary product to $\star$ with the 2-cocycle $s\alpha$, and its star product $\star_s$, $s \in [0,1]$ (i.e. $\star_1=\star$ and $\star_0$ is the ordinary product). Due to the physical consistency (\ref {consistency}) for any values of $s$ the mapped anomaly $\hat{\cal{A}}_{T,H}$ defines an integral cohomology class in $H^{2n}_{\star_s}(X,\mathbb{Z})$, i.e. $\hat \int \hat{\cal{A}}_{T,H} \in \mathbb{Z}$. Actually, $\hat \int \hat{\cal{A}}_{T,H}$ is fixed and is equal to $ \varepsilon_H\zeta_T \int_X \emph{\emph{ch}}_d(E)$, for $s=0$, since in this case we obtain \cite {bertlman, weinberg};\footnote{Actually, the translation-invariant noncommutative Yang-Mills theories with ordinary action receive an overal $(-)$ sign for anomaly. This is due to their original definition of currents which accounts an extra commutation between Grassmannian fields $\psi$ and $\overline \psi$ within the path-integral formalism. However, for the inverse action the sign of the anomaly coincides with that in the commutative case. To see this one should note that replacing $A^a_\mu \to -A^a_\mu$, as appeared in the interaction term, will not touch the triangle diagram amplitude in the commutative case. Thus, the gauge invariance determines the whole form of the chiral anomaly in commutative Yang-Mills theories with the inverse action. It remains only the case of the adjoint action. Obviously, as mentioned above, the adjoint action in the commutative case leads to free fields with no interaction and no anomaly, providing $\zeta_C=0$.}
\begin{equation} \label {proof-th-6}
\partial_\mu j^\mu_{T,H}=\frac{{\varepsilon_H\zeta_T}}{{n!}} \frac{{1}}{{(4\pi)^n}}~ Tr\{F_{\mu_1\nu_1}\cdots F_{\mu_n \nu_n} \} ~\epsilon^{\mu_1\nu_1 \cdots \mu_n \nu_n}=\varepsilon_H \zeta_T \emph{\emph{ch}}_n(E)~,
\end{equation}

\noindent wherein $\zeta_A=1$, $\zeta_B=-1$ and $\zeta_C=0$. Therefore, according to \textbf{Theorem 2} $\hat{\cal{A}}_{T,H}$ is cohomologous to $\varepsilon_H\zeta_T \hat{\emph{\emph{ch}}}_n(E)$. Thus, $\hat{\omega}_{T,H}=\hat {\cal{A}}_{T,H}-\varepsilon_H \zeta_T \check{\emph{\emph{ch}}}_d(E) \in \frak{X}_{2n}$ have to be a gauge covariant exact form. That is $\omega_{T,H}=\zeta^{-1}(\hat \omega_{T,H})$ is also a gauge covariant exact differential form on $X$. Hence, $\omega_{T,H}$ vanishes over $U$, as a dense trivialization open set of $E\to X$. On the other hand, since it belongs to the image of $d_X$ it contains no singularity at $X-U$. Therefore, $\omega_{T,H}=0$ on $X$. This proves the theorem. \textbf{Q.E.D} \\

\par Actually, upon \textbf{Theorem 3} the anomalous aspects of the chiral currents imply a trivial topology for the type C theory in the setting of $\star$-cohomology. It actually confirms our previous statement that the adjoint action results in decoupling of the gauge and the matter fields for the commutative case. Moreover, the above theorem provides some exact solutions for the general translation-invariant noncommutative Yang-Mills theories via a homotopic approach in the $\star$-cohomology classes. Thus, without performing any loop calculation due to the perturbative methods the formulations of the singlet and the chiral Abelian anomalies of the general translation-invariant noncommutative $U(N)$-Yang-Mills theory in four-dimensional spacetime, with the ordinary action, are derived as;
$$  \partial_\mu j^\mu_5+i[A^b_\mu,j^{b\mu}_5]_\star= -\frac{{1}}{{16 \pi^2}}~Tr\{ F_{\star \mu\nu} \star F_{\star \sigma \lambda}\} ~\epsilon^{\mu \nu \sigma \lambda}=-\frac{{1}}{{32 \pi^2}}~F_{\star \mu\nu}^a \star F_{\star \sigma \lambda}^a ~\epsilon^{\mu \nu \sigma \lambda}~,$$
 $$  \partial_\mu j^\mu_H+i[A^b_\mu,j^{b\mu}_H]_\star= \varepsilon_H \frac{{1}}{{32 \pi^2}}~Tr\{ F_{\star \mu\nu} \star F_{\star \sigma \lambda}\}  ~\epsilon^{\mu \nu \sigma \lambda}=\varepsilon_H \frac{{1}}{{64 \pi^2}}~F_{\star \mu\nu}^a \star F_{\star \sigma \lambda}^a  ~\epsilon^{\mu \nu \sigma \lambda}~.$$

\noindent Moreover, the axial and the chiral anomalies of four-dimensional noncommutative QED with the translation-invariant star product $\star$, for the inverse action, are given by the following exact solutions:
\begin{equation} \label {ncqed}
\mathcal{D}_\mu j^\mu_5= \frac{{1}}{{16 \pi^2}} F_{\star \mu\nu} \star F_{\star \sigma \lambda} ~\epsilon^{\mu \nu \sigma \lambda}~,~~~~~\mathcal{D}_\mu j^\mu_H=- \varepsilon_H \frac{{1}}{{32 \pi^2}} F_{\star \mu\nu} \star F_{\star \sigma \lambda} ~\epsilon^{\mu \nu \sigma \lambda}~.
\end{equation}

\noindent with $\mathcal{D}_\mu=\partial_\mu+[A_\mu,~]_\star$.

\par Actually, in order to extract these solutions by means of the perturbative methods one has to consider the quadrangle and the pentagon diagrams, although with finite contributions, to derive the unique gauge covariant (equivariant) result for the anomalous behavior of the axial current \cite {bertlman}. Doing loop calculation for general translation-invariant noncommutative star product $\star$, which in usual has no simplifying symmetry for handling the overal contributions of planar/non-planar diagrams (such as the symmetric structures which exist for that of the Moyal case\footnote{See for example \cite {gross-wulkenhar1, gross-wulkenhar2, ghasemkhani-varshovi} for the importance of such symmetries to organize the contribution of planar/non-planar loops with the Moyal overal phase factors. In \cite {gross-wulkenhar1} and \cite {gross-wulkenhar2} the authors has considered the renormalization of two- and four-dimensional Moyal noncommutative $\phi^4$ theory with regards to the intrinsic symmetry of the Moyal product. In \cite{ghasemkhani-varshovi} also the authors studied the quantum effects to photon mass in two-dimensional Moyal noncommutative QED, wherein the symmetric property of the Moyal product has a serious role to find an exact solution. Such simplifying symmetries are totally removed for a general translation-invariant noncommutative star product.}), would be a quite complicated procedure. This in fact, clarifies the significant role of the given cohomological exact solutions.

\par Indeed, proving the anomaly freeness of the adjoint action, $\mathcal{A}_{C,H}=0$, in the perturbation approach has similar complications due to the lack of mentioned symmetries, while it only needs to deal with the triangle diagram \cite {bertlman}.


\par
\section{Consistent Anomalies in Noncommutative Yang-Mills}
\setcounter{equation}{0}

\par To extend our formulations for general translation-invariant noncommutative Yang-Mills theories to consistent anomalies we should give a short review on this topic.\footnote{Actually, studying the consistent anomalies in noncommutative Yang-Mills theories with general translation-invariant star products follows just the same way as for that of the Moyal case \cite {varshovi-first}.} To formulate such theories within a more powerful framework one has to extend the superalgebra $\frak{X}$ in order to include the gauge transformations. To do this let us first consider $\frak{g}=\frak{su}(N)$, for some $N$, and set $\frak{X}'=\frak{X}\otimes \frak{g}$. Similarly, extend the domain of $d$ to $\frak X'$. Hence, $(\frak X',d)$ is also a differential graded algebra with $\frak X'=\oplus_{k=0}^{2n}\frak X'_{k}$, for $\frak X'_k=\frak X_k \otimes \frak{g}$. As it was cleared above $\frak X'_0$ is a unital $*$-algebra with $1=\hat 1 \mathbb{I}$ and $(\hat \omega^a t^a)^*=\hat \omega^* {t^a}^{\dag}$, wherein we assume a summation on the group parameter $a$. Also, an element of $\frak X'$, say $\hat a$, is said to be (anti-) Hermitian if we have; $\hat a^*=\hat a$ ($\hat a^*=-\hat a$).

\par The space of Hermitian elements of $\frak X'_1$, which is shown with $\mathcal A$, is an Affine space and is considered as the space of connection forms. Also, the space of anti-Hermitian elements of $\frak X'_0$, shown with $\tilde {\frak g}$, is a Lie algebra and is assumed as the space of infinitesimal gauge transformations. Thus, a connection form is generally denoted by $\hat A=\hat A^a_\mu t^a~dx^\mu$, whereas an infinitesimal gauge transformation has a general form of $\hat \alpha =-i\hat \alpha^a t^a$. By definition, the infinitesimal transformation of $\hat{A}$ with respect to $\hat{\alpha}$ is;
\begin{equation} \label {gauge transformation1}
         {}_{{\hat{\alpha}}}\delta \hat{A}:={}_{{\hat{\alpha}}}\delta \hat{A}^a_\mu t^a~dx^{\mu}=-id\hat{\alpha}+[\hat{\alpha},\hat{A}]~,
\end{equation}

\noindent of which we find;
\begin{equation} \label {nctransformation}
{}_\alpha \delta A_\mu^a:= \partial_\mu \alpha^a-\frac{{1}}{{2}}f^{abc} \{\alpha^b,A^c_\mu\}_\star -\frac{{i}}{{2}}c^{bca} [\alpha^b,A_\mu^c]_\star~,
\end{equation}

\noindent for any type $T=A,B,C$. The Yang-Mills action (\ref {ymaction}) is invariant with respect to infinitesimal gauge transformation. The infinitesimal gauge transformation of spinors is accordingly given by;
\begin{equation} \label {gauge transformation2}
\begin{array}{*{20}{c}}
    \bold{Type~A:}~~~~~~~~~~{}_{\hat \alpha} \delta \hat{\psi}:=\hat{\alpha}\hat{\psi}=i\hat{\alpha}^a t^a\hat{\psi} ~,~~~~~~~~~~~~ \\
    \bold{Type~B:}~~~~~~~~~~{}_{\hat \alpha} \delta \hat{\psi}:=\hat{\psi}\hat{\alpha}^{*}=-i t^a\hat{\psi}\hat{\alpha}^a ~,~~~~~~~~ \\
   \bold{Type~C:}~~~~~~~~~~{}_{\hat \alpha} \delta \hat{\psi}:=\hat{\alpha}\hat{\psi}+\hat{\psi}\hat{\alpha}^*=i[t^a\hat{\psi},\hat{\alpha}^a] ~. \\
\end{array}
\end{equation}

\noindent It is seen that the matter action $\mathcal{S}^T_{int}=\mathcal{S}_D+\mathcal{S}^T_{int}$ remains invariant via (\ref {gauge transformation1}) and (\ref{gauge transformation2}) due to (\ref {dirac}) and (\ref {matter}).

\par The Lie algebra $\tilde{\frak{g}}$ generates a Lie group $\tilde{G}$ by elements $e^{\hat{\alpha}}=\sum_n \frac{{1}}{{n!}}\hat{\alpha}^n$, for $\hat{\alpha} \in \tilde{\frak{g}}$. Due to (\ref {gauge transformation1}) the Lie group $\tilde{G}$ acts on $\mathcal A$ from right; $\hat{A} \to \hat{A}^{\hat g}=\hat g^{-1}d\hat g + \hat{g}^{-1} \hat{A} \hat g$. The exterior derivative on $\tilde G$, $\tilde d$, defines the BRST operator. According to the Maurer-Cartan formula and the gauge transformation (\ref {gauge transformation1}) we readily see that;
\begin{equation} \label {brst-first}
\tilde d \hat c = i \hat c^2~,~~~~~~~~~~\tilde d \hat A=-d\hat c +i[\hat c,\hat A]~,~~~~~~~~~~\tilde d \hat F_\star =i [\hat c, \hat F_\star]~,
\end{equation}

\noindent for $\hat c=\hat c^a t^a$ the Maurer-Cartan form on $\tilde G$, the so called ghost field. Here, $[\hat a,\hat b]=  \hat a \hat b -(-)^{|a||b|}~\hat b \hat a$ is the super-commutator. Obviously, we obtain $ \tilde d^2=0$, while the supersymmetric relation $ \tilde d d =- d \tilde d$ is considered accordingly. In fact:
\begin{align} \label {m-brst-c}
\tilde d \hat c^a = -\frac{{1}}{{4}}~ f^{abc}~ \{\hat c^b, \hat c^c\}-\frac{{i}}{{4}} ~c^{bca}~[\hat c^b,\hat c^c]~,~~~~~~~~~~~~\\
\label {m-brst-A}
\tilde d \hat A^a_\mu=-\partial_\mu \hat c^a-\frac{{1 }}{{2}}~f^{abc}~ \{ \hat c^b,\hat A^c_\mu \} -\frac{{i }}{{2}}~c^{bca}~ [\hat c^b,\hat A^c_\mu]~,
\end{align}

\noindent for $\{\hat a,\hat b\}=\hat a.\hat b+(-)^{|\hat a| |\hat b|} ~ \hat b. \hat a$ the super anti-commutator due to the $\mathbb{Z}_2$-grading of $\frak{X}'$ with $|\hat c^a|=1$. From (\ref {m-brst-c}) and (\ref {m-brst-A}) one can simply derive the $\star$-version formulations as;
\begin{align} \label {ncbrst-c}
\tilde d  c^a = -\frac{{1}}{{4}}~ f^{abc}~ \{ c^b,  c^c\}_\star -\frac{{i}}{{4}} ~c^{bca}~[ c^b, c^c]_\star~,~~~~~~~~~~~~\\
\label {ncbrst-A}
\tilde d A^a_\mu=-\partial_\mu c^a-\frac{{1 }}{{2}}~f^{abc}~ \{ c^b, A^c_\mu \}_\star -\frac{{i }}{{2}}~c^{bca}~ [ c^b, A^c_\mu]_\star~,
\end{align}

\noindent for $\{a,b\}_\star=a\star b+(-)^{|a| |b|} ~ b\star a$ and $[a,b]_\star=a\star b-(-)^{|a||b|}~b\star a$, the super $\star$(-anti)-commutators.

\par Now, let us define $\tilde X=\mathbb{R}^2 \times X$, wherein $\mathbb{R}^2$ is considered to contain commutative coordinates $x^{-1}$ and $x^{-2}$. Since $\tilde X$ deformation retracts to $X$ then we find; $H^{2n+2}_{dR}(\tilde X,\mathbb{C})=H^{2n+1}_{dR}(\tilde{X},\mathbb{C})=0$. This, due to \textbf{Proposition 1} results in $H^{2n+2}_\star (\tilde{X},\mathbb{C})=H^{2n+1}_\star(\tilde X,\mathbb{C})=0$, which actually leads to triviality of $H^{2n+2}_\star (\tilde{X},\mathbb{Z})=H^{2n+1}_\star(\tilde X,\mathbb{Z})=0$. Here, the elements of $H^{2n+2}_\star (\tilde{X},\mathbb{Z})$ and $H^{2n+1}_\star(\tilde X,\mathbb{Z})$ are defined upon the generalization of the integration structure (\ref {integral}). Therefore, the third type Chern character $\check{\emph{\emph{ch}}}_{n+1}(E)$ in $2n+2$ dimensions (and the corresponding singlet anomaly) is an exact form. That is, there exists a $2n+1$-form, say $\hat \Pi_{2n+1}^0$, so that; $d \hat {\Pi}_{2n+1}^0(\hat A)=\check{\emph{\emph{ch}}}_{n+1}(E)$. Here the superscript $(0)$ indicates the number of the ghosts included, i.e. the so called ghost number. But, actually, to derive the Stora-Zumino chain of descent equations for general translation-invariant noncommutative star product $\star$ we need a more strict formulation of $\hat{\mathcal{A}}_{Singlet,T}(\hat A)$ via a modification of $Tr$.

\par Let $\frak B$ be a superalgebra generated by $B=\{\hat A^a t^a, d\hat A^a t^a, \hat c^a t^a, d\hat c^at^a\}$ with $|\hat A^a|=|\hat c^a|=1$ and $|d\hat A^a|=|d\hat c^a|=0$. Thus, any element of $\frak B$ is a linear combination of elements $\hat a_1. \cdots .\hat a_k $ where each $\hat a_i$ belongs to $B$. It is a well-defined decomposition of elements in $\frak B$, so called the gauge-ghost decomposition. Applying the gauge-ghost decomposition we define the supersymmetric trace $STr$ with
\begin{equation} \label {str}
   STr\{ \hat a_1 \cdots \hat a_p  \}=\frac{{ 1 }}{{p}}~ \sum_{k=1}^{p}~(-1)^{\sigma_k}~Tr\{ \hat a_k \cdots \hat a_p. \hat a_1 \cdots \hat a_{k-1} \}~, 
\end{equation}

\noindent where $\sigma_k=|\hat a_1 \cdots \hat a_{k-1}||\hat a_k \cdots \hat a_p|$. It is easily seen that $STr$ vanishes on super-commutators. Moreover, if $\hat \omega \in \frak B$ is given by $Tr$, one defines $\hat \omega^{Sym} \in \frak B$ by replacing $Tr$ with $STr$ accordingly. \\

\textbf{Lemma 2;} \emph{Assume that $\hat \omega \in \frak B$ is defined by $Tr$ operator and $\hat \omega^{Sym}$ is its supersymmetric form. Then we have;}
\par \textbf{a)} \emph{If $\hat \omega$ is gauge covariant then $\hat \omega^{Sym}$ is gauge invariant. Consequently; $\tilde d \hat \omega^{Sym}=0$.}
\par \textbf{b)} \emph{If $d\hat \omega =0$, then; $d\hat \omega^{Sym}=0$, but the inverse is not true in general.}
\par \textbf{c)} \emph{Integration of $\hat \omega$ and $\hat \omega^{Sym}$ coincides, i.e.; $\hat \int \hat \omega = \hat \int \hat \omega ^{Sym}$.}
\par \textbf{d)} \emph{If $\hat \omega$ is a form of degree $2n$ or $2n+1$ or $2n+2$, then $\hat \omega$ and $\hat \omega^{Sym}$ are cohomologous.} \\

\par \textbf{Proof;} Actually, the main key of the proof of \textbf{(a)} is to note that any gauge transformation is in fact a super-commutator. This leads to \textbf{(a)} as a conclusion. Moreover, it is obvious from definition (\ref{str}) that if $\hat \omega$ is closed, then ${\hat \omega}^{Sym}$ would be closed too. However, the converse is not true. For instance, $STr\{ \hat F_{\star \mu_1\nu_1} \cdots \hat F_{\star \mu_{k}\nu_{k}} \}$ is closed, but $dTr\{ \hat F_{\star \mu_1\nu_1} \cdots \hat F_{\star \mu_{k}\nu_{k}} \} \ne 0$ in general. This proves \textbf{(b)}. On the other hand, since the integration process is a trace operator for the star product $\star$, then the integral of any summand in the right hand side of (\ref{str}) is equal to $\hat \int \hat{\omega}$. Then \textbf{(c)} follows by the definition. Finally, \textbf{(d)} is the immediate result of $H^{2n+2}_\star (\tilde{X},\mathbb{C})=H^{2n+1}_\star(\tilde X,\mathbb{C})=0$ and \textbf{(c)}.\footnote{However, for more details of the proof and similar results for the Moyal product see \cite {varshovi-consist, varshovi-first}.} \textbf{Q.E.D} \\

\par Due to \textbf{Lemma 2} $\check{\emph{\emph{ch}}}_{n}^{Sym}(E)$ and $\check{\emph{\emph{ch}}}_{n+1}^{Sym}(E)$ both are gauge invariant closed forms with the integral cohomology classes equal to those of $\check{\emph{\emph{ch}}}_{n}(E)$ and $\check{\emph{\emph{ch}}}_{n+1}(E)$. So let us to replace the singlet anomaly $\hat {\mathcal{A}}_{Singlet,T}(\hat A)$ with $\hat {\mathcal{A}}_{Singlet,T}^{Sym}(\hat A)=d\Omega^{0}_{2n+1,T}(\hat A)$ to work out a matrix base descent equation. We have;
$$\hat {\mathcal{A}}_{Singlet,T}^{Sym}(\hat A)=- 2\zeta_T \check{\emph{\emph{ch}}}_{n+1}^{Sym}(E):=-\frac{{2\zeta_T}}{{(n+1)!(4\pi)^{n+1}}}~STr\{ \hat F_{\star \mu_1\nu_1} \cdots \hat F_{\star \mu_{n+1}\nu_{n+1}} \}~\epsilon^{\mu_1\nu_1 \cdots \mu_{n+1}\nu_{n+1}}~d^{2n+2}x~.$$

\noindent However, by \textbf{Lemma 2 (a)} one finds; $\tilde d \hat {\mathcal{A}}_{Singlet,T}=0$. Thus, due to the anti-commutativity of $d$ and $\tilde d$ we readily have; $d \tilde d \hat {\Omega}_{2n+1,T}^0(\hat A)=0$, which together with the triviality of $H^{2n+1}_\star(\tilde X,\mathbb{C})$ shows that $\tilde d \hat {\Omega}_{2n+1,T}^0(\hat A)$ is itself an exact form. Hence, there exists a differential form, say $ \hat {\Omega}_{2n,T}^1 (\hat A) $, so that $d\hat {\Omega}_{2n,T}^1(\hat A)=-\tilde d \hat {\Omega}_{2n+1,T}^0(\hat A)$. Actually, we accomplished a matrix base formulation of the Stora-Zumino chain of descent equations:
\begin{align} \label {descent-equation}
\begin{array}{*{20}{c}}
 \hat{\mathcal{A}}_{Singlet,T}^{Sym}(\hat A)-d\hat \Omega_{2n+1,T}^0(\hat A)=0~,\\
\tilde d \hat \Omega^0_{2n+1,T}(\hat A)+d\hat \Omega_{2n,T}^1(\hat A)=0~,~~~~\\
~~\tilde d \hat \Omega^1_{2n,T}(\hat A)+d\hat \Omega_{2n-1,T}^2(\hat A)=0~,~~~~~~  \\
\end{array}
\end{align}

\noindent where the last equation is worked out locally due to the classical Poincare lemma. The differential forms $\widehat {CS}_{2n+1}(\hat A):=-\frac{{1}}{{2\zeta_T}}~\hat \Omega_{2n+1,T}^0(\hat A)$, $\hat G_{2n,T}(\hat A):=-\pi \hat \Omega_{2n,T}^1(\hat A)$ and $\widehat {Sc}_{2n-1,T}(\hat A):=-\pi \hat \Omega_{2n-1,T}^2(\hat A)$ are conventionally known as the $2n+1$-dimensional Chern-Simons form, the consistent anomaly and the consistent Schwinger term.\footnote{Actually, the overal factor for defining the consistent anomaly and the consistent Schwinger term is $-2\pi$, but here due to our convention we should insert an extra $\frac{{1}}{{2}}$ to transfer to the chiral case of $L-$ or $R-$ currents. See \cite {bertlman, nakahara}.} Direct calculation shows that \cite {varshovi-first};
\begin{align}
\label {CS-4}
\hat \Omega_{5,T}^0(\hat A)= - \frac{{ \zeta_T}}{{24 \pi^3}} ~STr \{ \hat A (d\hat A)^2 +\frac{{3}}{{2}} \hat A^3 d\hat A + \frac{{3}}{{5}} \hat A ^5 \}~,~~~~ \\
\label {consistent-4}
\hat \Omega_{4,T}^1(\hat A)=- \frac{{\zeta_T}}{{24 \pi^3}} ~STr \{ \hat c ~d(\hat A d \hat A + \frac{{1}}{{2}} \hat A^3)\} ~, ~~~~~~~~~~~~~~\\
\label {sch-4}
\hat \Omega_{3,T}^2(\hat A)= \frac{{\zeta_T}}{{48 \pi^3}}~STr \{ (\hat c^2 \hat A+\hat c \hat A \hat c + \hat A \hat c^2 ) d\hat A + \hat c^2 \hat A^3 \} ~.~
\end{align}

\noindent and;
\begin{align}
\label {CS-2}
\hat \Omega_{3,T}^0(\hat A)= - \frac{{ \zeta_T}}{{4 \pi^2}} ~STr \{ \hat A d\hat A +\frac{{2}}{{3}} \hat A^3  \}~,~~~~
\hat \Omega_{2,T}^1(\hat A)= - \frac{{\zeta_T}}{{4 \pi^2}} ~STr \{ \hat c  d \hat A \} ~,
\end{align}

\noindent without well-defined $\hat \Omega_{1,T}^2(\hat A)$.\footnote{Actually, the chain of descent equations could be continued as far as the spacetime dimension of $\hat \Omega^{k}_{2n+1-k}(\hat A)$, i.e. $2n+1-k$, becomes (at least) 2 to maintain the well-defined structure of the translation-invariant noncommutative star product $\star$. Nevertheless, a blind calculation leads to $\hat \Omega_{1,T}^2(\hat A)= \frac{{\zeta_T}}{{4 \pi^2}}~STr \{ \hat c^2 \hat A \}$ which we avoid to report it here.} The given formulas (\ref {CS-4})-(\ref {CS-2}) are in fact the exact derivations via the considerable abilities of the introduced differential graded algebras and the $\star$-cohomology. \\

\par \textbf{Theorem 4;} \emph{Assume the general translation-invariant noncommutative star product $\star$ over $X$. Then, replacing $d_X$ by $d$ for simplicity we find;}\footnote{See \cite {varshovi-first} for comparable results for Moyal star product.}
\par \textbf{a)} \emph{In four-dimensional noncommutative Yang-Mills theories with the star product $\star$ we have:}\footnote{For the Moyal noncommutative Chern-Simons form for QED see \cite {ghasemkhani}. The consistent anomaly is studied by several aouthors. See for example \cite {bonora, brandt, varshovi-consist}, where \cite {brandt} uses the Seiberg-Witten map for deriving an explicit solution for the Moyal product. But \cite {varshovi-consist} solves the problem for general translation-invariant noncommutative star products by means of a matrix formulation. For the noncommutative consistent Shwinger term see also \cite {varshovi0, langman, varshovi-consist}.}
\begin{align*}
{CS}_{5~\star}(A)=\frac{{1}}{{144\pi^3}}~Tr \{  \{ A, (d A)_\star^2\}_\star + dA \star A \star dA + \frac{{9}}{{8}} \{ A_\star^3 , d A\}_\star +\frac{{9}}{{8}} \{A \star dA \star A, A\}_\star + \frac{{9}}{{5}} A_\star ^5 \}~,~~~~~~~~~~~~~\\
G_{4,T~\star} (A)=  \frac{{\zeta_T}}{{72 \pi^2}} ~Tr \{  \{ dA, c \star d A\}_\star + (dA)^2_\star \star c + \frac{{3}}{{8}}  \{ \{c,dA\}_\star , A^2_\star \}_\star + \frac{{3}}{{8}} \{\{A,c\}_\star, \{ dA, A \}_\star \}_\star \} ~,~~~~~~~~~~~~~~~\\
 Sc_{3,T ~\star}(A)=-\frac{{\zeta_T}}{{192 \pi^2}}~Tr \{ \{ \{c^2_\star,A\}_\star + c\star A\star c,dA\}_\star+\{\{A\star dA, c\}_\star,c\}_\star+\{c\star dA \star c, A \}_\star ~~~~~~~~~~~~~~~~~~~~~\\ 
  +\frac{{\zeta_T}}{{240 \pi^2}} ~Tr \{\{c,c\star A_\star^3 \}_\star + \{A, A\star c^2_\star \star A \}_\star +A^3_\star \star c^2_\star \}~,~~~~~~~~~~~~~~~~~~~~~~~~~~~~~~~~~~~~~~~~~~~~~~~~~~
\end{align*}

\noindent \emph{wherein $a^n_\star$ means $n$ times multiplication of $a$ with the star product $\star$.}
\par \textbf{b)} \emph{In two-dimensional noncommutative Yang-Mills theories with the star product $\star$ we have:}
\begin{align} 
  CS_{3~\star}(A)=\frac{{1}}{{16 \pi^2}}~Tr \{\{A,dA\}_\star +\frac{{4}}{{3}} A_\star^3 \}~,  \\
  G_{2,T~\star}(A)=  \frac{{\zeta_T}}{{8 \pi}} ~Tr \{ \{ c, dA \}_\star \} ~,~~~~~~~~~~~~~
\end{align}

\noindent \emph{wtih no well-defined consistent Schwinger term}.\footnote{As mentioned before $Sc_{1,T~\star}(A)$ is not well-defined. But, however, a blind algebraic calculation shows that $Sc_{1,T~\star}(A)=- \frac{{\zeta_T}}{{12 \pi}} ~Tr \{ \{ c^2_\star, A\}_\star + c\star A \star c \} \}$.}  \\

\par It can be seen that the complicated formulas of \textbf{Theorem 4} give rise to the familiar formulations for the consistent anomaly and Schwinger term with the ordinary or the Moyal product via the corresponding integrations \cite{bertlman, varshovi-consist}. For instance, we easily see from \textbf{Theorem 4 (a)} that;
$$\int_X G_{4,T~\star} (A)=\frac{{\zeta_T}}{{24 \pi^2}} ~\int_X Tr \{ c \star d( A \star d A + \frac{{1}}{{2}} A_\star^3)\}~,$$

\noindent which has the familiar formula up to the $\star$ product.

\par The superalgebra $\frak B$ also has a natural grading due to the ghost number of the elements. Thus, $(\frak B, \tilde d)$ is itself a differential graded algebra and produces a cohomology theory, so called the $\star$-BRST cohomology, which its groups are shown as $H^k_{\star -BRST}(\tilde G, \mathbb{C})$. On the other hand, the Stora-Zumino equations show that $\widehat{CS}(A):=\hat \int CS_{2n+1}(A)$, $\widehat G(A):=\hat \int G_{2n,T}(A)$, and $\widehat{Sc}(A):=\hat \int Sc_{2n-1,T}(A)$ all are BRST closed forms, and thus represent cohomology classes in $H^0_{\star -BRST}(\tilde G,\mathbb{C})$, $H^1_{\star -BRST}(\tilde G,\mathbb{C})$, and $H^2_{\star -BRST}(\tilde G, \mathbb{C})$ respectively.\footnote{As indicated above, the integral structure $\hat \int$ is generalized to hyper-spacetime $\mathbb{R}\times X$ and the space $X/\mathbb{R}$ in a concrete manner.} Moreover, upon the derivation procedure of the exact solutions in \textbf{Theorem 4} due to the corresponding Chern-Weil theory and the Chern-Simons transgression, these cohomology classes are independent of the connection form $A$. Consequently, $\widehat{CS}$, $\widehat G$ and $\widehat{Sc}$ are topological invariants of the corresponding noncommutative Yang-Mills theory. In principle, these BRST differential forms belong to the integral classes of $H^k_{\star -BRST}(\tilde G,\mathbb{Z})$, for $k=0,1,2$, and thus their integration over $k$-dimensional submanifolds in $\tilde G$ reveal some topological indices of the theory. This topic is studied by means of the noncommutative index theorem in a separate work \cite {varshovi-forth}.


\par
\section{Summary and Conclusions}
\setcounter{equation}{0}

\par In this article we introduced a new cohomology theory, the so called $\star$-cohomology, with groups $H_\star^k(X,\mathbb{C})$, $k \geq 0$, to study the geometry/topology of general translation-invariant noncommutative Yang-Mills theories over the spacetime manifold $X$. It was shown that the $\star$-cohomology theory provides an algebraic breakthrough between the de Rham and the cyclic (co)homology theories, so that it enables us to translate the cyclic (co)homology classes to that of the de Rham even for the case of noncommutative algebras such as that of the noncommutative quantum fields. In principle, the $\star$-cohomology classifies operator valued differential forms on the commutative part of $X$ with extra dimensions on its noncommutative part. This gives rise to a homotopy formula to transfer from translation-invariant noncommutative Yang-Mills theories to those of the commutative case. Hence, it is a cohomological formulation of the Seiberg-Witten map for general translation-invariant noncommutative star products.

\par Actually, it was proven that the machinery of the $\star$-cohomology provides a simple algebro-geometric framework to study the topological aspects of any modified version of general translation-invariant noncommutative gauge theories. More precisely, the $\star$-cohomology was shown to be a suited and an economical noncommutative version of the de Rham cohomology (from the algebro-geometric insights) for the noncommutative quantum fields. Therefore, any new formulation of translation-invariant noncommutative Yang-Mills theory (provided by adding some exotic counter/interaction terms which vanish in transferring from the case of noncommutative star products to the ordinary commutative fields) could also be investigated via the topological features and the underlying geometry in the setting of the $\star$-cohomology.

\par Continuing our studies three types of noncommutative Yang-Mills theories were introduced with the ordinary (type A), the inverse (type B), and the adjoint (type C) actions of the gauge fields on the spinors. Employing the Chern-Weil theory via the framework of $\star$-cohomology formulation, we introduced an integral cohomology class in $H_\star^k(X,\mathbb{Z})$, the so called third type Chern character, which reveals the geometric and topological structures of the chiral anomalies of the mentioned three types A, B, and C by means of the integral classes of the $\star$-cohomology. The exact solutions for the Abelian anomalies were derived and it was established that the type C is anomaly free for the chiral current. Finally, the noncommutative versions of the Chern-Simons form, the consistent anomaly and the consistent Schwinger term for general translation-invariant noncommutative star products were worked out by transgression the third type Chern character.

\par Actually, it was shown that the $\star$-cohomology also provides a well-defined machinery to work out the anomalous structures, and hence the non-renormalizable features, of the theory. Therefore, the $\star$-cohomology could be considered as an appropriate algebro-geometric method for studying the (non-) renormalizability of the modified versions and the different types (as investigated in this paper) of general translation-invariant noncommutative Yang-Mills theories. For instance, as it was established in the paper, the type C theory admits no non-renormalizable structure from the chiral anomaly. Finally, the $\star$-BRST cohomology groups were defined accordingly and the integral cohomology classes of the Chern-Simons form, the consistent anomaly and the consistent Schwinger term were described via their geometric formalisms.

\section{Acknowledgments}
\par The author says his special gratitude to S. Ziaee who was the main motivation for appearing this article. Moreover, the author wishes to dedicate this work to the memory of Abbas Kiarostami, the creator of "Khaneye Dust Kojast", the Iranian filmmaker, writer, poet, photographer and painter, for all his elegant contributions to the Iranian art and culture. Also, it should be noted that this research was in part supported by a grant from IPM (No.99810422).










\end{document}